\documentclass[acmtog, screen, nonacm]{acmart}
\AtBeginDocument{%
  }

\setcopyright{acmlicensed}
\copyrightyear{2018}
\acmYear{2018}
\acmDOI{XXXXXXX.XXXXXXX}
\acmConference[ ]{ }{ }{ }
\acmISBN{ }

\usepackage{multirow}
\usepackage{longtable}
\usepackage{subfigure}
\usepackage{subcaption}
\usepackage{graphicx}
\usepackage{makecell}
\usepackage{graphicx}

\begin{document}

\title{From Adaptation to Intelligence: A Systematic Review of Data, Strategies, and Impact in Personalized VR}

\author{Tangyao Li}
\email{tli724@connect.hkust-gz.edu.cn}
\orcid{0009-0008-6157-9294}
\author{Yitong Zhu}
\email{yzhu162@connect.hkust-gz.edu.cn}
\orcid{0009-0007-5717-3390}
\author{Hai-Ning Liang}
\email{hainingliang@hkust-gz.edu.cn}
\orcid{0000-0003-3600-8955}
\author{Yuyang Wang}
\authornote{Corresponding author}
\email{yuyangwang@hkust-gz.edu.cn}
\orcid{0000-0003-0242-8935}
\affiliation{%
  \institution{The Hong Kong University of Science and Technology (Guangzhou)}
  \city{Guangzhou}
  \state{Guangdong}
  \country{China}
}


\begin{abstract}

As virtual reality (VR) systems advance, they are increasingly expected to adapt intelligently to individual users' states, abilities, and preferences. While prior research has examined user-state sensing and adaptive interaction design in VR, existing reviews typically address these aspects in isolation. In this paper, we examine the growing body of research on personalization in VR, with a particular focus on how user data collected during immersion is used to drive adaptive strategies that tailor the experience and enhance engagement, performance, or other specific goals. We synthesize findings from studies that employ adaptive techniques across diverse application domains and summarize a five-stage conceptual framework that unifies adaptive mechanisms across domains. Our analysis reveals emerging trends, including the integration of multimodal sensors, the transition from purely reactive to hybrid adaptation systems, and the adoption of artificial intelligence approaches. Finally, we identify key challenges related to data, modeling, and evaluation, and outline future research directions toward more effective and user-centered VR systems.

\end{abstract}






\maketitle

\section{Introduction} \label{sec:intro}

Virtual reality (VR) offers users an immersive experience by creating a virtual environment that leverages the combined power of hardware, software, tracking technology, and perceptual science \cite{ribeiro2023virtual,sun2018towards}. Through various immersive setups, especially with the use of head-mounted displays (HMDs), VR enables experiences that are difficult or impossible to achieve through traditional interfaces \cite{tewell2024review}. Advances in VR hardware, particularly HMDs and motion tracking technologies, have made immersive VR experiences more accessible and robust. The first HMD, ``The Ultimate Display'', was invented by Ivan Sutherland in the mid-1960s \cite{sutherland1965ultimate}. The landscape of VR has undergone a significant transformation over the last decade. The release of the consumer versions of Oculus Rift and HTC Vive in 2016 made VR more accessible to a much wider audience, but the high cost and technical limitations are drawbacks \cite{renganayagalu2021effectiveness}. Oculus Go, the first standalone VR headset, was released in 2018 at a more affordable price. Later, the Oculus (now Meta) Quest series further improved immersion and interactivity by supporting six degrees of freedom tracking through inside-out sensing \cite{helou2023virtual}. Recent advancements include the 2024 release of Apple Vision Pro and PICO 4 Ultra. Nowadays, multiple standalone VR HMDs are available on the consumer market and have been widely adopted in domains such as gaming, training, healthcare, and education~\cite{baghaei2021anxiety,wang2023movement,Xu2023acceptance,Elsholz2025VRtaxonomy,Qawqzeh2025training}.

However, users differ substantially in their skills, preferences, and cognitive capacities, making one-size-fits-all VR experiences insufficient. To address these differences, many VR systems attempt to tailor the experience to individual users to improve user experience (UX). Personalized interaction, also referred to as adaptive VR or personalized VR, encompasses a range of strategies \cite{pardini2022role,correia2024adaptive}. These may include, but are not limited to, customizing virtual environments to enhance immersion, adjusting visual elements for better visual clarity, and adapting the pace or presentation of content to support individual learning styles. In general, there are three main stages of an adaptive VR system: data acquisition, real-time user modeling, and adaptive interactive mechanisms \cite{baker2022adaptive}.

Recent advances in sensing technologies and artificial intelligence (AI) have significantly expanded the ways in which user information in the virtual environment can be collected and interpreted \cite{pavel2025patchfusionvr,ribeiro2023virtual,Wang2025AI4VR}. Multimodal data, such as behavioral and physiological signals, enable systems to infer user state in real-time, which overcomes the issue of delayed feedback from using a post-experiment questionnaire \cite{zhang2014detection,wood2021virtual}. Machine learning (ML) algorithms further support this process by modeling complex relationships between user data and system responses, allowing adaptive behaviors to be learned from data rather than being manually specified \cite{halkiopoulos2025role}. Despite rapid progress in VR and human-computer interaction (HCI) research, existing research on personalized VR remains fragmented across technological approaches, adaptation mechanisms, and evaluation practices.

With the rapid development of VR-related hardware and technology, a comprehensive overview is needed to consolidate knowledge, identify methodological trends, and highlight gaps across different technical and application domains. Thus, we synthesize recent empirical studies on adaptive VR systems that utilize real-time or offline user data to modify the experience. This systematic review focuses on answering the following research questions:

\textbf{RQ1}: What are the current technological approaches for implementing personalized VR?

\textbf{RQ2}: How is the adaptive logic in VR systems evolving, particularly with the rise of AI?

\textbf{RQ3}: How do current personalization strategies impact the user experience?

\subsection{Scope}

This survey focuses on personalization and adaptivity in VR systems that employ physiological, behavioral, or contextual information to tailor UX dynamically. In this work, VR refers to the fully immersive virtual experience powered by computer-generated three-dimensional (3D) environments through devices such as HMDs. This excludes augmented and mixed reality systems that blend virtual and physical worlds together within a shared perceptual space. Section \ref{sec:criteria} provides a more detailed explanation of the inclusion and exclusion criteria. Systems that focus solely on content creation or non-adaptive VR experiences are excluded. The objective of this survey is to synthesize existing methods, modalities, and adaptive mechanisms that enable personalized VR experiences, and to identify emerging trends and key challenges.

\subsection{Contributions}

Our survey provides a systematic synthesis of recent research on personalized and adaptive VR. We organize adaptive VR systems using a five-stage conceptual framework and provide an integrative analysis of how personalization manifests across distinct domains. By examining AI-driven approaches, biosignal utilization, and user-centered evaluation, this work identifies trends, methodological gaps, and challenges that shape future directions in adaptive VR research.

We organize this paper as follows: Section \ref{sec:related} discusses previous reviews on adaptive VR systems across different domains. Section \ref{sec:methods} outlines the methodology used to identify and select the reviewed papers. Section \ref{sec:results} presents findings guided by three research questions. Section \ref{sec:discussion} provides an in-depth discussion of identified emerging trends. Section \ref{sec:future} provides suggestions about future research directions regarding data, model, and evaluation levels. Finally, we conclude this paper in Section \ref{sec:conclusion}.

\section{Related Surveys} \label{sec:related}

To contextualize our contribution, we position our review alongside a selection of related works that examine VR personalization, adaptive systems, or UX from various disciplinary and application-specific angles. Wood et al. \cite{wood2021virtual} conducted a comprehensive analysis of physiological sensing (e.g., heart rate, skin responses, gaze, brain, and muscle activity) in VR. They identified a growing trend in using these measures as user input devices for customized and adaptive user experiences. Tewell and Ranasinghe \cite{tewell2024review} focused on olfactory technology for VR, classifying it by presentation approach, delivery method, and application area. In terms of immersion experience, Gon\c{c}alves et al. \cite{gonccalves2022systematic} analyzed the impact of different levels of realism on UX in immersive virtual experiences. They categorized realism into subjective and objective aspects and examined its influence on user performance, presence, and physiological responses.

Bergsnev and S\'anchez Laws \cite{bergsnev2022personalizing} explored the personalization of VR for research and treatment of fear-related disorders, focusing on self-report measures and manipulations of contextual factors. They highlighted the methodological challenges and the need for greater interactivity in VR procedures for therapeutic applications. Similarly, Maddalon et al. \cite{maddalon2024exploring} reviewed adaptive VR systems used in interventions for children with autism spectrum disorder. They categorized adaptive strategies by level switching, feedback, and temporal dimensions (real-time vs. deferred), and analyzed the underlying adaptive engines (ML vs. non-ML) and signals (explicit vs. implicit).

Additionally, several papers reviewed the application of adaptation in education and training \cite{checa2020review}. Scott et al. \cite{scott2016adaptive} offered a review to emphasize learner models, instructional strategies, and adaptive mechanisms in educational VR. Similarly, Marougkas et al. \cite{marougkas2024personalized} examined personalization strategies and gamification techniques in immersive VR for educational purposes. They highlighted the insufficient incorporation of adaptive learning content in earlier studies. Richlan et al. \cite{richlan2023virtual} examined VR intervention studies in sports contexts, focusing on performance enhancement, neurocognitive mechanisms (e.g., visual search behavior and imagery), and methodological aspects, including adaptive training difficulty. Also, they discussed the transferability and generalizability of skills learned in VR to the real world. Aguilar Reyes et al. \cite{aguilar2023design} focused on the design and evaluation of a prototype adaptive VR training system for military pilot training, leveraging trainee performance measures, adaptive logic, and adaptive variables. Lucas-P\'erez et al. \cite{lucas2024personalising} presented a conceptual framework for integrating AI into immersive VR training systems to create adaptive learning environments.

While these studies collectively advance our understanding of VR personalization, they tend to focus on either user-state sensing, adaptive interaction design, or specific applications in isolation, thereby limiting insights into how different components jointly shape the VR experience. Therefore, this review addresses this gap by systematically examining the relationships between user-state sensing and adaptive mechanisms, highlighting how they work together to influence UX in VR, and what technological limitations need to be addressed in future studies.

\section{Methodology} \label{sec:methods}

By following the PRISMA guidelines \cite{moher2009preferred}, we conducted a comprehensive survey of published research on personalized user behavior in VR over the past ten years.

\subsection{Article Collections}

\begin{figure}
  \centering
  \includegraphics[width=\linewidth]{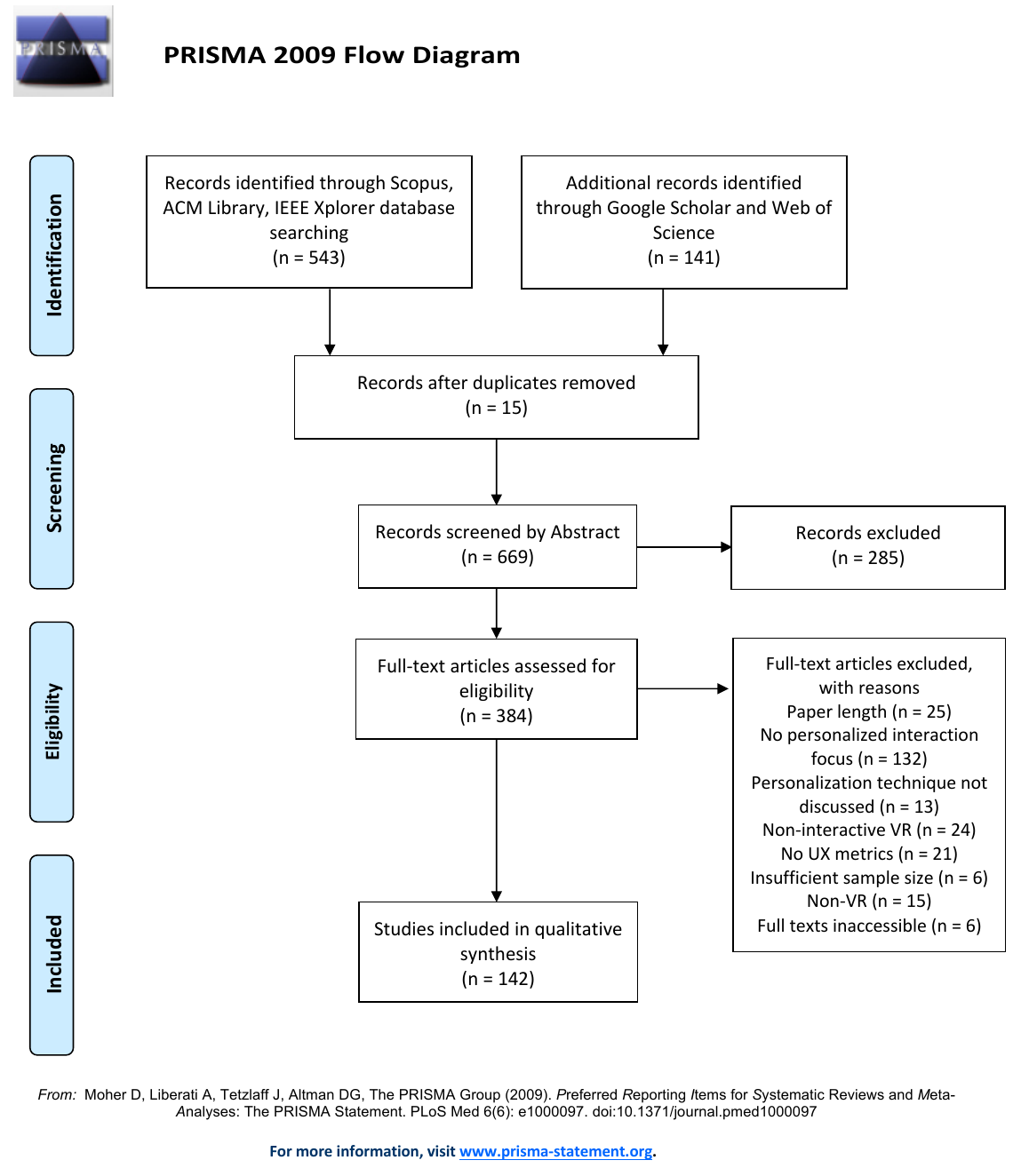}
  \caption{PRISMA flow diagram \cite{moher2009preferred} detailing the study selection process for the systematic review.}
  \label{fig:prisma}
\end{figure}

A comprehensive literature search was conducted across multiple academic databases, including Scopus, ACM Digital Library, IEEE Xplore, Web of Science, and Google Scholar, to identify relevant studies for this review. Over 142 studies were included. The search process was carried out in several rounds, with an initial retrieval in late 2024 and an updated retrieval in late 2025. After each round, we applied a two-step screening workflow: (1) title and abstract screening, followed by (2) full-text review, as reflected by the systematic literature review process in Figure \ref{fig:prisma}. The primary screening and selection process was conducted by one author to maintain consistency, with input from co-authors during stages such as refining search keywords and discussing inclusion criteria. A combination of keywords and Boolean operators was used to retrieve studies focusing on personalized user behavior in VR. The search query included terms related to personalization (e.g., personalized user interaction, adaptive user interaction, user personalization, customized experience), VR technology (e.g., virtual reality), and user behavior metrics (e.g., user behavior, UX, user engagement, user immersion). This keyword-based retrieval was complemented by author-based searches and citation tracing to identify additional relevant studies. We limited the final analyzed set to peer-reviewed journal and conference publications as a baseline quality filter.

\subsection{Eligibility Criteria} \label{sec:criteria}

We established a comprehensive set of inclusion and exclusion criteria to guarantee the relevance and quality of the studies incorporated in this systematic review. These criteria were specifically designed to focus on research directly advancing our understanding of personalized user behavior within VR environments.

\subsubsection{Inclusion Criteria}

Studies were included in this review based on four main criteria, as summarized in Table \ref{tab:criteria}. First, eligible studies focused on personalized interaction techniques within VR environments, aiming to explore how VR systems or applications can effectively adapt to the unique preferences and behaviors of individual users. Second, the scope of inclusion extended across all domains where VR is applied, including but not limited to gaming, education, healthcare, training simulations, and social VR, thereby enabling a comprehensive examination of personalization strategies across diverse fields. Third, only peer-reviewed journal articles and conference papers were considered, ensuring that the studies included met standards of academic rigor and credibility. Lastly, to reflect the rapid evolution of VR technologies, the review was limited to studies published in the last 10 years (from 2014 to 2025), capturing contemporary developments and trends in personalized VR interaction.

\begin{table}
    \centering
    \small
    \caption{Inclusion and exclusion criteria}
    \label{tab:criteria}
    \begin{tabular}{l}
        \toprule
        Inclusion criteria \\
        \midrule
        Research conducted in any field where VR is used \\
        Studies focusing on personalized user interaction techniques in VR \\
        Published in peer-reviewed journals or conferences \\
        Studies published in the last 10 years \\
        \toprule
        Exclusion criteria \\
        \midrule
        Publications shorter than 4 pages \\
        Studies not focused on personalization techniques \\
        Studies that omitted discussion of personalization techniques \\
        Studies that examine non-interactive VR experiences \\
        Studies failing to report UX or performance metrics \\
        Insufficient sample size \\
        Content not related to VR \\
        Publications with inaccessible full texts \\
        \bottomrule
    \end{tabular}
\end{table}

\subsubsection{Exclusion Criteria}

Several exclusion criteria were applied to maintain the quality and relevance of the selected literature, as detailed in Table \ref{tab:criteria}. First, studies were excluded if they were short papers lacking sufficient methodological detail, such as experimental design or data collection processes, or if they presented preliminary ideas without empirical validation. Second, research that did not focus explicitly on personalized user interaction within VR environments was also excluded to preserve the thematic scope of the review. Additionally, studies that implemented personalization but failed to describe their methods were excluded, as this hinders reproducibility and limits interpretability. Such papers include those that omit information on algorithmic or personalization techniques and those presenting non-interactive VR experiences, such as users watching passive 360-degree videos without real-time user engagement, were excluded as they fell outside the review's focus on interaction. Similarly, studies that did not report user engagement or performance metrics were also omitted, as such data are essential for evaluating the impact of personalization strategies. Further, studies with insufficient sample sizes (fewer than five participants) or lacking reported user performance data were excluded to ensure the reliability of findings. Finally, works not directly related to VR, such as those primarily focused on augmented reality, mixed reality, or general HCI studies, were excluded, as were inaccessible papers for which full content could not be reviewed.

\section{Results} \label{sec:results}

\begin{figure*}
    \centering
    \includegraphics[width=0.9\linewidth]{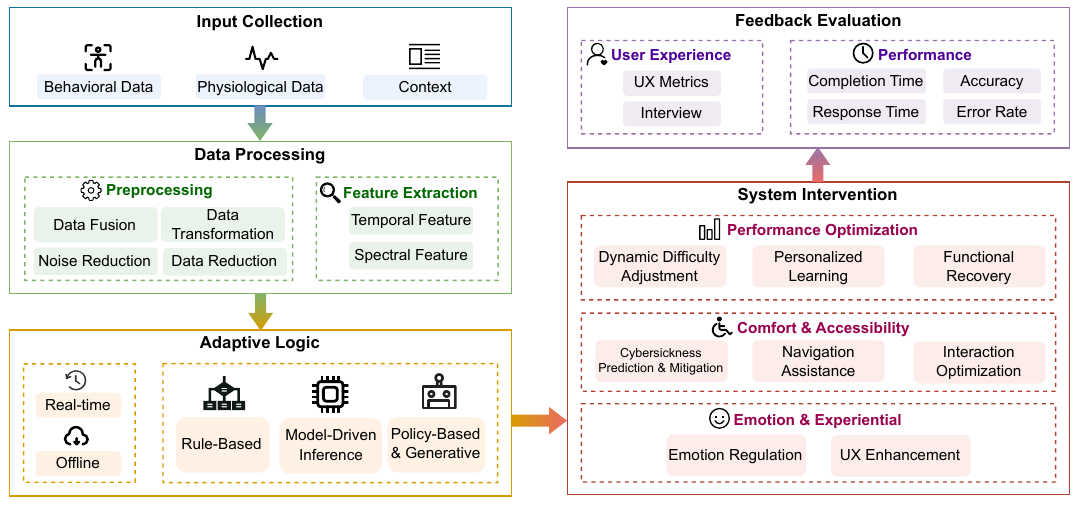}
    \caption{General pipeline for adaptive VR systems for personalized user experiences, including (1) Input Collection, (2) Data Processing, (3) Adaptive Logic, (4) System Intervention, and (5) Feedback Evaluation.}
    \label{fig:pipeline}
\end{figure*}

To systematically synthesize and analyze research progress in this field, we propose a five-stage adaptive framework (see Figure \ref{fig:pipeline}). This framework deconstructs personalized systems into a continuous process: (1) Input Collection, (2) Data Processing, (3) Adaptive Logic, (4) System Intervention, and (5) Feedback Evaluation. The empirical results in this section and subsequent in-depth discussion (Section \ref{sec:discussion}) will follow this framework, aiming to clearly deconstruct the technical chain from data collection (Section \ref{sec:rq1}) and adaptive algorithms (Section \ref{sec:rq2}) to UX impact (Section \ref{sec:rq3}).

\subsection{RQ1: Input Modalities} \label{sec:rq1}

Our review indicates that leveraging multimodal data to capture user states has become the mainstream paradigm in personalized VR research. To gain a deeper understanding of what modalities are used in current research, we analyze different modalities by category (see Table \ref{tab:biosignal}). Rather than relying on a single data source, recent systems increasingly combine multiple input modalities to capture users' internal states and interaction patterns more comprehensively. This trend is driven by the need to improve the accuracy, robustness, and generalizability of user state estimation under dynamic VR conditions.

First is the use of physiological signals. With the development of physiological sensors, data acquisition devices have moved from bulky laboratory equipment to wearable and consumer-grade devices, making them more suitable for real-world deployments. Among the reviewed studies, electrodermal (EDA), electroencephalography (EEG), and eye-tracking are the most commonly used modalities as they provide relatively direct access to users' internal cognitive and affective states. EDA is most commonly employed for cybersickness detection and estimation because it measures arousal-related responses, such as in \cite{islam2021cybersense,tasnim2024investigating,kundu2023litevr,sameri2024physiology,li2023deep,wang2023modeling,zhu2025towards,hadadi2022prediction}. Several studies also leverage EDA to infer broader user states, such as emotion classification \cite{caldas2024breaking}, workload estimation \cite{dubovi2022cognitive,chiossi2023adapting,chiossi2024optimizing}, and engagement measurement \cite{bian2019design}. In addition, EEG records brain activity in real-time by non-invasively placing electrodes on the scalp \cite{choi2023neural}, often used for user state classification due to its ability to capture neuro-related activities \cite{qin2024data,jung2024personalized,liang2023eeg,bekele2016multimodal,benlamine2021bargain,feick2024predicting,lopes2025generative}. For example, De Lima et al. \cite{de2022adaptive} used EEG and EDA as indicators of users' emotional engagement and stress levels to adjust their horror game's difficulty. Similarly, ocular-related features (eye-tracking and pupillometry) play a central role in assessing visual attention, cognitive states, and certain physiological responses, as shown in \cite{eckert2021cognitive,grootjen2024your,alghofaili2019lost,wang2023prediction}.

Behavioral and interaction-based signals refer to modalities that are obtained passively from user behavior, such as eye-tracking, posture, interaction, and motion data. Eye-tracking occupies a special position at the boundary between physiological and behavioral sensing, as it captures both motor behavior and cognitive processes such as attention and fatigue \cite{lahiri2012design}. For example, Zhao and Cheng \cite{zhao2023adaptive} proposed an adaptive navigation assistant that uses pupil center-corneal reflection techniques combined with ML to predict user navigation intentions. In addition, gaze can be used for alternative interaction input to enable hands-free interaction \cite{monteiro2021hands,lu2021itext}. Motion and posture data are frequently used to infer activity level, offering insights into user strategies and task engagement \cite{besga2025task}. Unlike biosignals, which require an additional acquisition device, a key advantage of these signals is that they can be obtained directly from HMDs \cite{grootjen2024your,zhao2023adaptive} or other external devices \cite{lahiri2012design} non-invasively.

Beyond the two types of signals mentioned above, researchers also use text, acoustic, and video inputs. These modalities appear less frequently than physiological or interaction-based signals, but they are mainly used in systems that support direct communication between users and virtual environments or agents, such as \cite{wan2024building,liu2024classmeta,bellucci2025immersive}. By converting speech to text, such systems enable natural language understanding and intent inference. For example, Hu et al. \cite{hu2025gesprompt} allow users to combine gestures and voice commands to modify objects in a virtual environment. In Ashby et al.'s work \cite{ashby2023personalized}, players could communicate with non-playable characters (NPCs) in a fantasy-based role-playing game, supporting more natural conversational interaction. Siddiqui et al. \cite{siddiqui2023manifest} and Monteiro et al.~\cite{Monteiro2024audience} analyzed users' speech and body language during a public speaking training to assess their speech and presentation quality. In these systems, automated speech recognition models, such as Whisper and Whisper-Tiny, are used in \cite{bellucci2025immersive,siddiqui2023manifest,wan2024building}, while text-to-speech models (e.g., ElevenLabs and Azure Text to Speech) are used in \cite{bellucci2025immersive,liu2024classmeta} to generate spoken responses. For video input, Sexton et al. \cite{sexton2021automatic} feed 360$^{\circ}$ videos into their video prediction model to generate scene-based olfactory cues delivered through an olfaction device. Zhu et al. \cite{zhu2025towards} used VR videos to predict users' cybersickness levels during immersion.

In short, current research predominantly employs multimodal fusion approaches to provide richer data for feature learning in the subsequent stage. On the other hand, problems like data fusion and alignment, computational cost, and user comfort arise. We provide a more in-depth discussion of these problems in Section \ref{sec:dis1}.

\begin{table*}[htbp]
    \centering
    \caption{Summary of physiological and behavioral metrics employed in empirical VR studies, categorized by measurement modality and primary use cases.}
    \begin{tabular}{llllll}
    \toprule
    Signal & Primary Use & References & Signal & Primary Use & References \\
    \midrule
    
    \multirow{3}{*}{EEG} & Cognitive load & \cite{dey2019exploration,chiossi2025designing,chiossi2024optimizing,tremmel2019estimating} & \multirow{3}{*}{EMG/sEMG} & User state & \cite{i2018toward} \\
    & User state & \cite{qin2024data,jung2024personalized,liang2023eeg,bekele2016multimodal,benlamine2021bargain,feick2024predicting,lopes2025generative} & & Cognitive load & \cite{reidy2020facial} \\
    & Cybersickness & \cite{uyan2024cdms,sameri2024physiology} & & Muscle activity & \cite{dash2019design} \\
    \midrule

    \multirow{4}{*}{EDA} & User state & \cite{jung2024personalized,hadadi2024machine,kritikos2021personalized,caldas2024breaking,ranasinghe2018season} & \multirow{4}{*}{Ocular} & User state & \cite{zhao2023adaptive,fominykh2018conceptual,alghofaili2019lost,hadadi2024machine,feick2024predicting,zenner2024beyond} \\
     & Engagement & \cite{bian2019design} & & Engagement & \cite{lahiri2012design,porssut2021adapting,besga2025task} \\
     & Cybersickness & \cite{islam2021cybersense,tasnim2024investigating,kundu2023litevr,sameri2024physiology,li2023deep,wang2023modeling,zhu2025towards,hadadi2022prediction,qu2022bio} & & Reading speed & \cite{grootjen2024your} \\
     & Cognitive load & \cite{dubovi2022cognitive,chiossi2023adapting,chiossi2024optimizing} & & Cybersickness & \cite{tasnim2024investigating,kundu2023litevr,zhu2025towards} \\
    \midrule

    \multirow{2}{*}{HR} & User state & \cite{jung2024personalized,fominykh2018conceptual,hadadi2024machine,yu2025effects,hopf2019exploring,ranasinghe2018season} & \multirow{2}{*}{ECG} & User state & \cite{caldas2024breaking,i2018toward,lopes2025generative} \\
    & Cybersickness & \cite{islam2021cybersense,tasnim2024investigating,kundu2023litevr,hadadi2022prediction} & & Cognitive load & \cite{chiossi2024optimizing,chiossi2023adapting} \\
    \midrule

    \multirow{2}{*}{PPG} & User state & \cite{jung2024personalized} & \multirow{2}{*}{Blood pressure} & User state & \cite{yu2025effects} \\
    & Engagement & \cite{bian2019design} & & Cybersickness & \cite{sameri2024physiology,zhu2025towards,hadadi2024machine} \\
    \midrule

    \multirow{4}{*}{Posture} & Body position & \cite{ibanez2021using,tao2023embodying,boban2023unintentional} & \multirow{2}{*}{Respiration} & \multirow{2}{*}{Engagement} & \multirow{2}{*}{\cite{bian2019design}} \\
    & Head position & \cite{porssut2021adapting,zhu2025towards} & & & \\
    \cmidrule{4-6}
    & Hand position & \cite{xiong2024reach,lagos2022personalized} & \multirow{2}{*}{Temperature} & \multirow{2}{*}{Cybersickness} & \multirow{2}{*}{\cite{sameri2024physiology,hadadi2022prediction}} \\
    & Engagement & \cite{besga2025task} & & & \\
    
    \bottomrule
    \end{tabular}
    \label{tab:biosignal}
\end{table*}

\subsection{RQ2: Evolution of Adaptive Algorithms} \label{sec:rq2}

Once the necessary data are collected and processed to extract meaningful information, they are fed into adaptive algorithms for further computation. Adaptive logic describes the mechanisms by which an adaptive VR system determines when and how to modify system behavior in response to sensed user or contextual states. Although adaptive behavior can generally be abstracted as a conditional mapping from system and user states to adaptation actions, existing VR systems differ substantially in how these mappings are represented, authored, and executed. Based on the degree of explicitness and autonomy of the adaptive decision-making process, we can broadly categorize prior work into three classes of adaptive logic: explicit rule-based adaptation, model-driven inference with rule-based control, and policy-based or generative adaptation.

Explicit rule-based adaptation refers to systems in which adaptive decisions are guided by manually specified deterministic rules that directly map observed variables to predefined adaptation actions \cite{jokste2017rule}. In such systems, designers specify both the conditions under which adaptation occurs and the corresponding adaptive responses using logic structures, such as if-then rules \cite{lin2022study}, threshold-based conditions \cite{chiossi2023adapting}, and finite state machines \cite{benlamine2021bargain,valluripally2021rule,ojha2024dynamic}. For example, Lin et al. \cite{lin2022study} present a system that adjusts virtual learning content based on a computed value of learning style tendency, with eight presets corresponding to the learning styles (active, reflective, sensing, intuitive, visual, verbal, sequential, and global). Chiossi et al. \cite{chiossi2023adapting} proposed an adaptive system to adjust visual complexity by modifying the number of NPCs in the set, thereby controlling users' comfort, performance, and workload. This type of method is transparent and computationally efficient. Designers maintain full control over the adaptation logic. However, rule-based systems exhibit significant limitations. First, they require designers to manually define all possible scenarios. This process is feasible on a small scale, but becomes impractical for complex environments. Second, these systems lack generalization because they cannot infer user states or preferences beyond their explicit programming and predefined conditions \cite{jokste2017rule,zhao2016generation}, thus limiting their adaptability.

\begin{figure}
    \centering
    \includegraphics[width=\linewidth]{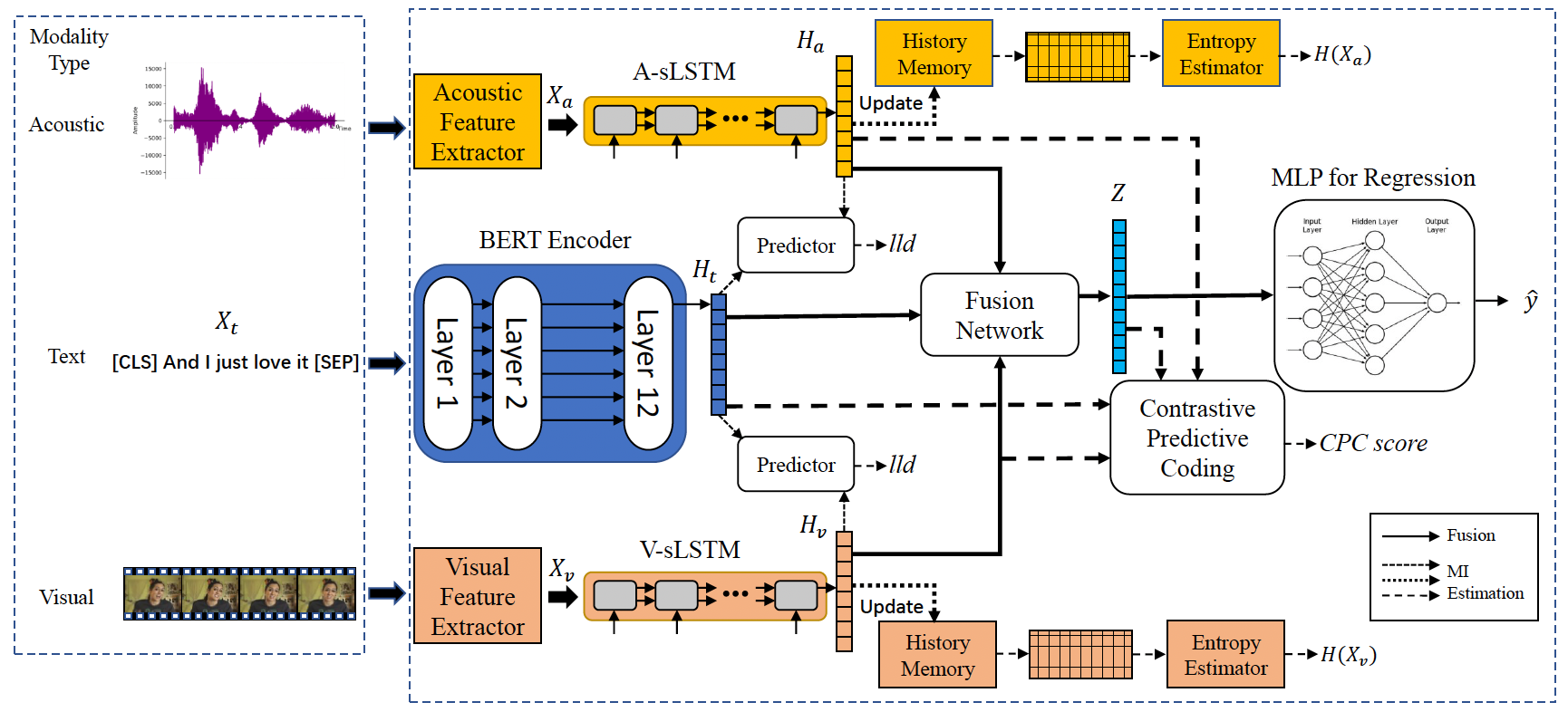}
    \caption{The structure of the MultiModal InfoMax model, which takes acoustic, text, and visual as inputs. Text input is encoded using BERT, while acoustic and visual sequences are processed by modality-specific LSTMs to generate unit-length representations $H_a$ and $H_v$. The fusion network integrates unimodal features and produces the fusion result $Z$. Contrastive predictive coding gives the CPC score that measures the mutual information between the multimodal fusion result $Z$. The multilayer perceptron (MLP) for regression is the final prediction module that gives the final prediction $\hat{y}$. From \cite{han2021improving}.}
    \label{fig:mmim}
\end{figure}

Model-driven inference systems employ machine learning (ML) models to estimate latent user states (e.g., emotion, engagement, cognitive load), while relying on predefined rules or mappings to select adaptive responses, see Table \ref{tab:technique}. In this category, machine learning is primarily used for perception and state estimation, not for adaptive decision generation. The inferred user state is subsequently mapped to adaptation actions using explicit logic defined by designers. Classical ML models, such as random forest \cite{feick2024predicting,bian2019design,raza2024optimized,hadadi2022prediction}, support vector machine \cite{asbee2023machine,hadadi2022prediction,raza2024optimized}, and k-nearest neighbors \cite{reidy2020facial,johnson2014knn}, are widely used for classification or prediction tasks from multimodal features. Additionally, various neural network architectures, including convolutional neural networks (CNNs) \cite{yu2025effects,lopes2025generative,sexton2021automatic,uyan2024cdms,zhu2025towards} and other deep neural networks \cite{wu2024gazefed,guo2018optimizing}, have been implemented. For example, in \cite{ibanez2021using}, a CNN is used to recognize emotions (happiness, horror, sadness, anger, surprise, and disgust) from head gestures in an adaptive VR music system to improve virtual presence. Some studies also employed ML models at earlier stages of the adaptive pipeline to process multimodal inputs and extract meaningful representations \cite{raza2024optimized,wu2024gazefed}. For instance, Han et al. \cite{han2021improving} used a pretrained language model BERT to encode textual inputs (see Figure \ref{fig:mmim}). The strengths of model-driven inference systems lie in the ability to incorporate multimodal data with improved robustness compared to handcrafted thresholds. Also, it is more scalable than purely rule-based systems while maintaining designer control over adaptation outcomes. Nevertheless, adaptive actions remain constrained to predefined options, limiting the capacity for creative adaptation strategies. In addition, errors in state inference may lead to mistakes in adaptation decisions.

Policy-based and generative adaptation refers to systems in which adaptive responses are learned, optimized, or generated dynamically, rather than selected from a predefined set of actions. The adaptive logic is encoded in a learned policy, objective function, or generative model. Designers specify high-level goals, constraints, or reward functions, while the system autonomously determines appropriate adaptation strategies at runtime. Policy-based adaptation typically employs reinforcement learning (RL), optimization, or planning techniques, where the system learns an adaptive policy that maximizes a predefined reward signal related to performance, engagement, or user experience \cite{ohnbar2018personalized,porssut2021adapting,yu2025effects}. For instance, Ohn-Bar et al. \cite{ohnbar2018personalized} proposed a model-based RL framework that could direct visually impaired users to their destination efficiently and safely via speech and haptic feedback. The system takes into account the mobility characteristics of blind users, such as moving speed and delayed responses to instructions, and incorporates a weighted expert module to adapt the dynamics model to new users. For generative adaptation, it leverages generative models to produce adaptive content, such as dynamically generated narratives, environments, tasks, or dialogues. Recent systems increasingly incorporate large language models (LLMs) to synthesize contextually appropriate content guided by prompts or constraints \cite{song2024developing,chheang2024towards,li2024exploring,min2024public,wan2024building,ashby2023personalized,gao2025pervrml}. Li et al. \cite{li2024exploring} used the LLM's ability to simulate natural conversation and develop a chatbot (powered by GPT-3.5 Turbo) with job-related communication training scenarios for autistic individuals. Similarly, Min and Jeong \cite{min2024public} introduced an LLM-generated (Claude-3) persona that could help presenters to practice a question-and-answer session. Their system featured a wide variety of questions based on the content of the presentation generated by LLMs and human experiment assistants. Based on the answer, virtual audiences will provide positive or negative feedback. This category of systems offers high flexibility and expressiveness, but often exhibits reduced interpretability and predictability. Also, there is the risk of generating unintended or undesirable content \cite{lopes2025generative}.

Across the reviewed literature, model-driven inference with rule-based control is the most widely used adaptive logic paradigm, reflecting the balance of personalization with interpretability and practicality. Fully generative adaptation approaches remain less common and are sometimes discussed as exploratory future work. Importantly, this trend should not be interpreted as a limitation of current techniques or that the ultimate goal is maximal personalization. The appropriateness of adaptive strategies depends on the application context, especially in domains where safety or controllability are key. As a result, adaptive logic mechanisms with constrained adaptations may be preferable in certain settings, despite offering less creative personalization. A more in-depth discussion of domain-specific adaptive requirements and their implications for adaptive logic design is provided in Section \ref{sec:dis2}.

\begin{table*}[htbp]
\footnotesize
\caption{Summary of ML-based techniques employed in empirical VR studies.}
\label{tab:technique}
\begin{tabular}{p{2.5cm} p{3.5cm} p{4cm} p{4cm} p{2cm}}
    \toprule
    Technique & Purpose & Strengths & Weaknesses & References \\
    \midrule
    Random Forest  & User behavioral analysis, physiological response classification & Handles high-dimensional data well; less prone to overfitting & Computationally intensive with a large number of trees; less interpretable & \cite{feick2024predicting,bian2019design,raza2024optimized,hadadi2022prediction} \\
    \midrule
    Long Short-Term Memory & Cybersickness prediction, physiological signal analysis, anomaly detection & Well-suited for time-series data; captures long-term dependencies; effective for complex sequence signals & Can be computationally expensive to train; architecture can be complex to tune & \cite{wang2019vr,alghofaili2019lost,yu2025effects,kundu2023litevr,tasnim2024investigating,miao2019deep,azim2025your,wan2024building} \\
    \midrule
    Convolutional Neural Network & Image recognition, physiological signal analysis, emotion recognition & Highly effective for image and spatial data analysis; scales with large datasets & Requires a large amount of data for effective training; can be computationally demanding & \cite{yu2025effects,lopes2025generative,sexton2021automatic,uyan2024cdms,zhu2025towards} \\
    \midrule
    Generative AI \& Large Language Models & NPC behavior and dialogue generation, personalized content creation, conversational virtual assistants & Enables dynamic personalized and context-aware interactions; reduces the need for manually creating context; highly flexible for conversational agents & Sometimes produce inaccurate responses; requires significant computational resources; raises ethical concerns regarding data privacy & \cite{song2024developing,chheang2024towards,li2024exploring,min2024public,wan2024building,ashby2023personalized,gao2025pervrml} \\
    \midrule
    Reinforcement Learning & Adaptive navigation, embodiment customization, feedback optimization & Effective for optimizing sequential decision-making; learns from interaction without explicit programming & Requires careful design of the reward function; requires a large number of iterations & \cite{ohnbar2018personalized,porssut2021adapting,yu2025effects} \\
    \midrule
    Artificial Neural Network & Emotion classification & Can learn complex non-linear relationships; effective for pattern recognition & Difficult to interpret; require a large amount of training data; prone to overfitting & \cite{de2022adaptive,ojha2024dynamic} \\
    \midrule
    Support Vector Machine & User performance classification, cybersickness prediction & Effective in high-dimensional spaces; memory efficient & Sensitive to data noise & \cite{asbee2023machine,hadadi2022prediction,raza2024optimized} \\
    \midrule
    k-Nearest Neighbors & Physiological signal classification, user performance classification & Simple to understand and implement; no training phase is needed & Computationally expensive with large datasets; sensitive to irrelevant features; performance depends on the choice of the `k' parameter & \cite{reidy2020facial,johnson2014knn} \\
    \midrule
    Generative Adversarial Network & Virtual environment generation & Effective for creating diverse content & Can produce a limited variety of outputs & \cite{yu2025effects} \\
    \midrule
    YOLO & Hand gesture tracking & Fast and accurate for real-time applications & Less effective at detecting small objects & \cite{karthick2023artificial} \\
    \midrule
    Deep Neural Networks & Gaze prediction & Can learn complex patterns & Requires large datasets; computationally expensive & \cite{wu2024gazefed,guo2018optimizing} \\
    \midrule
    Hidden Markov Model & User state prediction & Effective for modeling sequences and time-series data & Assumes the Markov property & \cite{brambilla2023tuning} \\
    \midrule
    Extreme Gradient Boosting & User behavior prediction & High performance & Prone to overfitting & \cite{zhao2023adaptive} \\
    \midrule
    Cascaded Regression Tree & Facial Feature point detection & Effective for regression tasks & Can be computationally intensive & \cite{chen20223d} \\
    \midrule
    BERT-ITE & User interaction data analysis & Specialized for sequential data & Requires large datasets; computationally expensive & \cite{li2023design} \\
    \midrule
    Fuzzy Logic & Cybersickness prediction & Can handle uncertainty and imprecision; interpretable & Tuning difficulties; difficult to define fuzzy rules & \cite{wang2021using} \\
    \midrule
    Multilayer Perceptron & Cybersickness detection, gesture recognition & Relatively simple to implement; scalable & Prone to overfitting; computationally intensive & \cite{kundu2023litevr,lopes2025generative} \\
    \bottomrule
\end{tabular}
\end{table*}

\subsection{RQ3: Effectiveness of Personalization Strategies} \label{sec:rq3}

This section synthesizes the effectiveness of adaptive strategies by organizing them into three categories, as shown in Table \ref{tab:strategy}. We examine: 1) Performance Optimization, which aims to optimize participants' performance (e.g., task efficiency and learning outcomes); 2) Comfort \& Accessibility, which seeks to minimize adverse effects like cybersickness and improve inclusivity; and 3) Emotion \& Experiential, which are designed to modulate affective states and enhance engagement. To evaluate UX, different questionnaires are used. Table \ref{tab:questionnaire} gives a summary of questionnaires used, categorized by measurement type.

\begin{table*}[]
    \centering
    \caption{Overview of personalization strategy types, functional roles, and representative studies.}
    \label{tab:strategy}
    \begin{tabular}{ll p{8cm}}
        \toprule
        Category & Functional Roles & References \\
        \midrule
        \multirow{3}{*}{Performance Optimization} & Difficulty Adjustment & \cite{de2022adaptive,marinho2024eyes,johnson2014knn,dey2019exploration} \\
        & Personalized Learning & \cite{lei2024design,song2024developing,chheang2024towards,obourdin2024unlocking,hadjipanayi2022arousing,jin2023development,li2024exploring,gao2025pervrml,min2024public,chan2023study,yu2025effects,lin2022study,hu2025gesprompt,liu2024classmeta} \\
        & Functional Recovery & \cite{dash2019design,reidy2020facial,jung2024personalized,zuki2024assessing,mahmoudi2021automated,lee2022towards,castillo2024design,tudor2015development,ahmad2020development,ip2018enhance,du2024lightsword,lagos2022personalized,zhai2021virtual,bouatrous2023new} \\
        \midrule
        \multirow{3}{*}{Comfort \& Accessibility} & Cybersickness Prediction and Mitigation & \cite{uyan2024cdms,islam2021cybersense,hadadi2024machine,wang2019vr,li2023deep,qu2022bio,tasnim2024investigating,kundu2023litevr,sameri2024physiology,hadadi2022prediction,valluripally2021rule,recenti2021toward,zhu2025towards,wang2023prediction,monteiro2021trajectory} \\
        & Navigation Assistance & \cite{zhao2023adaptive,argelaguet2014adaptive,rebelo2024adaptive,montano2019drift,alghofaili2019lost,montano2017navifields,ohnbar2018personalized,dong2021tailored} \\
        & Interaction Optimization & \cite{zenner2024beyond,bian2019design,lahiri2012design,pei2023embodied,wu2024fanpad,wu2024gazefed,cheng2023interactionadapt,bekele2016multimodal,feick2024predicting,chiossi2023adapting,delahaye2023avatar,wan2024building,tanaka2025detection,benda2024examining,cabrera2024indyvr,feick2023investigating,zeng2023perceptually,wang2022realitylens,zhao2019seeingvr,xiong2024reach,grootjen2024your,azim2025your,bellucci2025immersive,wei2023gaze,hu2025modeling} \\
        \midrule
        \multirow{2}{*}{Emotion \& Experiential} & Emotion Regulation & \cite{caldas2024breaking,fominykh2018conceptual,liang2023eeg,kritikos2021personalized,pardini2022role,i2018toward,brambilla2023tuning,chiossi2025designing,lopes2025generative,pizzoli2019user} \\
        & UX Enhancement & \cite{tao2023embodying,peer2019mitigating,waltemate2018impact,porssut2021adapting,sexton2021automatic,benlamine2021bargain,ojha2024dynamic,valmorisco2024enabling,raza2024optimized,ashby2023personalized,ranasinghe2018season,besga2025task} \\
        \bottomrule
    \end{tabular}
\end{table*}

Performance-oriented strategies aim to maximize user efficiency and accuracy in specific tasks through adaptive adjustment. These strategies dominate domains such as skill training, education, and functional recovery \cite{chheang2024towards,obourdin2024unlocking,jin2023development,dey2019exploration}. Researchers usually evaluate their effectiveness through objective performance metrics, such as task completion time, error rate, scores, and response time. For example, Dey et al. \cite{dey2019exploration} proposed a cognitively adaptive VR training system that uses real-time EEG measurements (specifically alpha activity) to optimize performance by dynamically adjusting task difficulty. The system continuously monitors the user's cognitive load (via EEG alpha activity) and increases the task difficulty if the cognitive load is too low and vice versa. The results suggest that the brain compensates for increased task demand without influencing response time. In addition, Du et al. \cite{du2024lightsword} incorporated a VR exergame to improve cognitive inhibition for older adults through adaptive difficulty levels. Two key factors for effective gamified training are motivation and satisfaction \cite{gizatdinova2022pigscape}. Therefore, they include music and rhythmic elements in the VR exergame to promote motivation and engagement. The recovery was demonstrated by improved performance on standardized cognitive tests (Stroop, Reverse Stroop, Go/NoGo tasks), with the benefits persisting for six months after the training concluded.

Comfort and accessibility strategies aim to reduce physical discomfort caused by VR use. These strategies also remove interaction barriers for users with special needs. Evidence of effectiveness mainly comes from subjective feedback (i.e., questionnaires, see Table \ref{tab:questionnaire}). Cybersickness (also referred to as VR sickness or simulator sickness in other literature) is a type of motion sickness experienced in VR and a major challenge to user comfort and enjoyment. According to the sensory conflict theory, it arises from a mismatch between motion cues in the virtual space and the actual motion in the physical world \cite{reason1975motion}. Users may experience different levels of discomfort due to individual differences, such as age, gender, and VR experience \cite{tian2022review}. Therefore, cybersickness is a factor that cannot be ignored in VR-related research. Studies often aim to predict and mitigate it to prevent its impact on VR tasks or overall UX. To measure the extent of discomfort it causes, researchers use surveys such as the Simulator Sickness Questionnaire (SSQ) \cite{kennedy1993simulator}, the Fast Motion Sickness Scale (FMS) \cite{keshavarz2011validating}, and the Motion Sickness Questionnaire (MSQ) \cite{graybiel1968diagnostic} to gather subjective measures. Objective monitoring often involves biosignals (e.g., EEG, EDA, HR) to detect physiological correlates of discomfort, then adjust VR content to prevent further cybersickness induction. Uyan and Celikcan \cite{uyan2024cdms} developed a real-time cybersickness detection system to adapt VR content factors (navigation speed, rendering parameters, and scene complexity) based on EEG signals. Additionally, physical data can predict cybersickness with high accuracy. For example, Li et al. \cite{li2023deep} achieved 97.8\% accuracy using only kinematic data. Moreover, non-invasive signals from HMDs combined with VR video could be used for cybersickness prediction \cite{zhu2025towards}.

Navigation and interaction assistance consist of features such as hand redirection \cite{aliza2024eye,feick2024predicting,azim2025your,tanaka2025detection,feick2023investigating}, drift correlation \cite{zenner2024beyond,montano2019drift}, and remapping \cite{benda2024examining,tao2023embodying} to ensure a continuous and fluent VR experience for users. Setting up a VR system usually requires a dedicated physical space. Users often need to move around during interaction and navigation. However, many users may have limited physical space available. As a result, continuous walking-based navigation can be interrupted or constrained. To address this issue, several studies have proposed solutions that adapt virtual environments to small physical spaces. For example, Rebelo et al. \cite{rebelo2024adaptive} introduced a technique to allow smooth VR navigation by customizing virtual environments to fit in small physical spaces. Likewise, Dong et al. \cite{dong2021tailored} present a perception-aware algorithm to restructure VR space so users can walk continuously without interruption in a small physical space. Furthermore, accessibility-focused interaction design makes VR more inclusive. To adapt to users' specific physical constraints, Zhao et al. \cite{zhao2019seeingvr} introduced a set of fourteen tools to offer visual and auditory enhancements to enable people with low vision (whose vision cannot be corrected with glasses or contact lenses) to use VR. For people with motor impairment, Xiong et al. \cite{xiong2024reach} proposed a hand redirection technique, Post Offset, for upper limb rehabilitation that improves patient motivation during rehabilitation exercises, and the redirection is hardly noticeable to patients. Pei et al. \cite{pei2023embodied} addressed the challenges of traveling for wheelchair users by enabling them to visit unfamiliar physical environments in virtual space. Their system adapts by allowing users to choose their avatar's physical appearance and input precise wheelchair measurements, which is identified to increase users' confidence. This design ensures that the virtual experience accurately mirrors their real-world capabilities and constraints for specific assessment tasks.

\begin{table*}[htbp]
    \centering
    \caption{Summary of commonly used questionnaires in the reviewed studies, categorized by measurement type.}
    \label{tab:questionnaire}
    \begin{tabular}{lll}
        \toprule
        Type & Questionnaire & References \\
        \midrule
        \multirow{3}{*}{Cybersickness}
        & Simulator Sickness Questionnaire \cite{kennedy1993simulator} & \cite{wang2019vr,dong2021tailored,uyan2024cdms,islam2021cybersense,zenner2024beyond,tanaka2025detection,sameri2024physiology,li2023deep,wang2023modeling,hadadi2022prediction,wang2021using} \\
        & Fast Motion Sickness Scale \cite{keshavarz2011validating} & \cite{hadadi2024machine,tasnim2024investigating,kundu2023litevr} \\
        & Motion Sickness Questionnaire \cite{graybiel1968diagnostic} & \cite{gao2025pervrml} \\
        \midrule
        
        \multirow{3}{*}{Presence}
        & Presence Questionnaire \cite{witmer1998measuring} & \cite{gironacci2022xr,dubovi2022cognitive,hopf2019exploring,ranasinghe2018season} \\
        & Igroup Presence Questionnaire \cite{schubert2001experience} & \cite{pazour2018virtual,wang2022realitylens,calvert2020impact,tanaka2025detection} \\
        & Slater-Usoh-Steed presence questionnaire \cite{slater1994depth} & \cite{ibanez2021using} \\
        \midrule
        
        \multirow{1}{*}{Usability}
        & System Usability Scale \cite{brooke1996sus} & \cite{chheang2024towards,lahiri2012design,zenner2024beyond,holzwarth2021towards,gao2025pervrml,rios2021user,hadjipanayi2022arousing,min2024public} \\
        \midrule
        
        \multirow{3}{*}{Emotion}
        & State-Trait Anxiety Inventory-Y \cite{spielberger1971state} & \cite{pardini2022role} \\
        & Self-Assessment Manikin \cite{bradley1994measuring} & \cite{liang2023eeg,waltemate2018impact,pardini2022role,brambilla2023tuning,i2018toward,lopes2025generative} \\
        & Positive and Negative Affect Scale \cite{watson1988development} & \cite{dubovi2022cognitive,lopes2025generative} \\
        \midrule
        
        \multirow{5}{*}{Experience}
        & Game Experience Questionnaire \cite{poels2007d3} & \cite{brambilla2023tuning,reidy2020facial,de2022adaptive,benlamine2021bargain,chiossi2025designing,chiossi2023adapting,ranasinghe2018season} \\
        & Virtual Reality Neuroscience Questionnaire \cite{kourtesis2019validation} & \cite{caldas2024breaking,castillo2024design} \\
        & Virtual Experience Questionnaire \cite{tcha2016proposition} & \cite{besga2025task,tanaka2025detection} \\
        & User Experience Questionnaire \cite{hinderks2017design} & \cite{gao2025pervrml,min2024public} \\
        & AttrakDiff-Short questionnaire & \cite{rios2021user} \\
        \midrule
        
        \multirow{2}{*}{Embodiment}
        & Embodiment Questionnaire \cite{peck2021avatar,gonzalez2018avatar} & \cite{tao2023embodying,zenner2024beyond,boban2023unintentional} \\
        & Alpha IVBO Scale \cite{roth2017alpha} & \cite{waltemate2018impact} \\
        \midrule
        
        \multirow{1}{*}{Immersion}
        & Immersive Tendencies Questionnaire \cite{singer1996presence} & \cite{gironacci2022xr,castillo2024design} \\
        \midrule   
        
        \multirow{1}{*}{Workload}
        & NASA Task Load Index \cite{hart2006nasa} & \cite{chheang2024towards,wu2024fanpad,holzwarth2021towards,chiossi2025designing,gao2025pervrml,chiossi2023adapting,grootjen2024your,wang2021enhanced} \\        
        \bottomrule   
    \end{tabular}
\end{table*}

\begin{figure}[htbp]
    \centering
        \subfigure[]{
        \label{fig:raza}
        \includegraphics[width=0.8\linewidth]{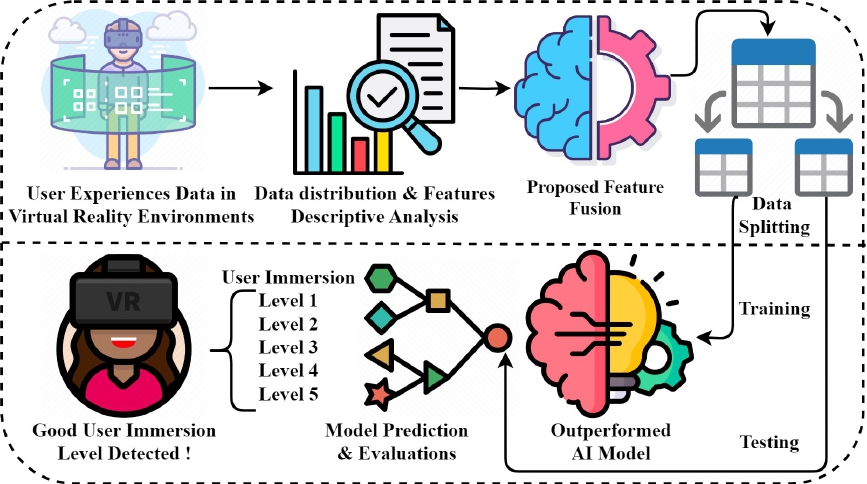}
    }
    \\
        \subfigure[]{
        \label{fig:sexton}
        \includegraphics[width=0.8\linewidth]{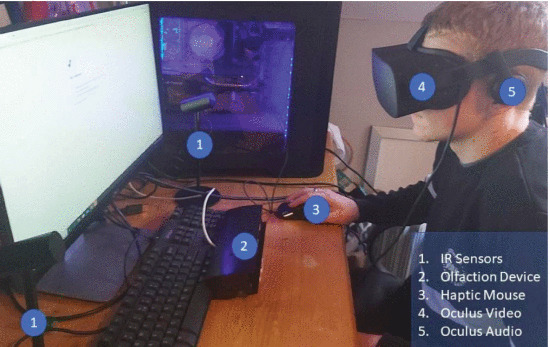}
    }
    \caption{(a) A framework for optimized VR design via immersion level detection. This system proposes a Polynomial Random Forest for feature generation and is capable of performing classification. From \cite{raza2024optimized}. (b) The hardware setup of a multiple-sensorial media platform that provides users with vision, audio, olfaction, and haptic effects according to immersive 360$^{\circ}$ videos. Video and audio are played on an Oculus Rift HMD, an Inhalio SBi4v2 olfaction dispenser provides olfaction, and a SteelSeries Rival 700 mouse gives haptic feedback. From \cite{sexton2021automatic}.}
\end{figure}

Emotion and experiential strategies focus on users' emotional states, psychological responses, and depth of immersion. These strategies commonly appear in psychotherapy and entertainment. Typically, self-report questionnaires often employ the Self-Assessment Manikin for measuring one's affective feeling \cite{bradley1994measuring}, the Game Experience Questionnaire for evaluating gaming experience \cite{ijsselsteijn2013game,poels2007d3}, and the NASA Task Load Index (NASA-TLX) for measuring perceived workload \cite{hart2006nasa}. With the estimated affective state from physiological signals, adaptive VR systems can dynamically modify virtual content to regulate or maintain emotions. For instance, Mahmoudi-Nejad et al. \cite{mahmoudi2021automated} proposed a personalized VR exposure therapy framework that estimates user stress from physiological data and applies content generation via RL to continuously adapt fear-inducing stimuli, ensuring stress remains within a therapeutic range. Similarly, Liang et al. \cite{liang2023eeg} developed an EEG-driven adaptive VR environment that automatically selects and inserts emotionally valenced stimuli into a predefined scene to promote relaxation and reduce negative emotions. More generally, Bermudez i Badia et al. \cite{i2018toward} introduced a software architecture for emotionally adaptive VR in which physiological signals drive procedural generation of audiovisual elements to reflect and influence users' affective states. Together, these works demonstrate how emotional strategies enable VR systems to respond to users' internal states in real-time, supporting emotion regulation and mental health interventions. On the other hand, we need to focus on several key UX aspects to improve users' overall UX. Depending on the research topic and aim, the targeted UX dimensions will be different. Since VR fundamentally seeks to evoke a sense of presence and immersion in a virtual environment, immersion is often treated as a broad and foundational UX objective in VR research. Raza et al. \cite{raza2024optimized} proposed a polynomial random forest technique for detecting user immersion levels through HR and EDA, see Figure \ref{fig:raza}. Additionally, Sexton et al. \cite{sexton2021automatic} proposed an automatic CNN-based enhancement system that could generate olfaction effects in immersive 360$^{\circ}$ videos, as shown in Figure \ref{fig:sexton}.

However, current evaluation paradigms show some limitations. Most studies rely on short-term performance metrics and immediate self-reports. This focus leaves long-term effects largely unexplored. Important questions remain unanswered. Performance optimization may encourage algorithm dependence. This dependence may weaken users' independent problem-solving skills. Comfort adjustments may also affect vestibular adaptation over long-term use. The long-term psychological effects of emotional regulation strategies also remain unclear.

\section{Discussion} \label{sec:discussion}

As personalized systems in VR transition to real-world deployment, a range of implementation challenges arise, spanning data (Section \ref{sec:dis1}), model (Section \ref{sec:dis2}), and evaluation (Section \ref{sec:dis3}) levels. While many studies demonstrate promising outcomes in controlled environments, transferring these results into daily application remains limited. In this section, we would like to discuss the causes of emerging trends and any potential challenges encountered. For clarity, we label the identified challenges as \textbf{CH1 – CH7}, which are later referenced in Section \ref{sec:future} when suggesting future research directions.

\subsection{Trends and Tensions in Multimodal Sensing} \label{sec:dis1}

Current personalized VR research exhibits a pronounced preference for multimodal data. The following sections will delve into the underlying reasons for this pattern (Section \ref{sec:biosignal}) and analyze the fundamental tension it reveals between technological feasibility and research practice (Section \ref{sec:practicality}).

\subsubsection{Predominance of Physiological Signals} \label{sec:biosignal}

As mentioned in Section \ref{sec:rq1}, EDA, EEG, and eye-tracking are the most commonly used sensing modalities in adaptive human-centered design. As a result, we would like to uncover factors that make them popular in HCI studies. The predominance of physiological signals in adaptive human-centered VR design can be attributed to both their theoretical relevance and practical feasibility. 

Across the reviewed studies, EDA and eye-tracking are the most consistently adopted modalities, particularly in applications requiring real-time adaptation, such as attention guidance and affect-aware feedback \cite{hadadi2022prediction,marinho2024eyes}. We observed that this trend emerged due to both theoretical relevance and its relatively favorable balance between signal quality and practical feasibility. Regarding theoretical relevance, EDA provides a stable proxy for arousal, and eye-tracking enables fine-grained attention estimation \cite{kosch2023survey}. Hence, these two modalities provide a way to measure affective state. In contrast, although EEG offers richer access to cognitive processes, its adoption is constrained in interactive VR scenarios due to motion sensitivity and setup complexity \cite{tauscher2019immersive}. Section \ref{sec:practicality} gives a more detailed discussion of this matter. Therefore, we can observe that researchers prefer modalities that are relatively easy to use, whereas modalities with higher theoretical fidelity are less often used under VR settings.

Moreover, hardware constraints further reinforce this imbalance. Firstly, the integration of eye-tracking into commercial-grade HMDs (e.g., HTC Vive Pro Eye, HTC Focus Vision, Meta Quest Pro, and Apple Vision Pro) has lowered the barrier to adopting eye measurement in VR systems. However, obtaining precise EEG data is relatively more difficult. EEG signals are acquired from the head region, so EEG devices usually adopt a head-mounted setup \cite{dey2019exploration}. As such, this type of setup may conflict with wearing VR HMDs simultaneously. Therefore, the predominance of EDA and eye-tracking could be due to the advantages of integrating into VR HMDs or simplifying acquisition devices. Despite the advantages in terms of hardware, we spot important limitations that are often underreported. For example, Sipatchin et al. \cite{sipatchin2021eye} and Wei et al. \cite{wei2023preliminary} demonstrate that eye-tracking accuracy is influenced by head movement and peripheral viewing conditions, suggesting that many adaptive systems may rely on signals whose reliability is context-dependent. This raises concerns about the generalizability of findings across different VR interaction scenarios. In contrast, EEG-based approaches tend to mitigate noise sensitivity through preprocessing and feature engineering \cite{tarrant2018virtual,feick2023investigating}. While this enables the extraction of meaningful indicators, it also leads to variations in the operational definitions of affective states across different studies. For example, different works emphasize distinct EEG frequency bands for similar constructs (e.g., workload, cybersickness, and engagement \cite{chiossi2024optimizing,chiossi2025designing,uyan2024cdms,benlamine2021bargain}), making cross-study comparisons difficult.

Taken together, these findings suggest a key constraint in current adaptive VR applications: systems tend to favor modalities that are relatively easy to deploy rather than those that provide the most comprehensive representation of user states. To address the limitations, many studies propose multimodal sensing as a way to balance precision, robustness, and usability \cite{zheng2022multimodal}. However, the review indicates that multimodal approaches are often motivated by compensating for individual sensor weaknesses rather than being guided by a principled understanding of how different modalities contribute complementary information. Although there are studies that investigate what combination of modalities gives the best prediction outcome, there is no universal agreement about the optimal combination. For example, Islam et al. \cite{islam2022towards} found that a cybersickness prediction model that combines eye-tracking with heart rate and EDA gives the best prediction result, while Jeong et al. \cite{jeong2023mac} showed that models that use EDA, VR video, and head and eye movement give promising results. As a result, there remains limited evidence on which modality combinations are most effective for specific adaptation scenarios.

\subsubsection{Balancing Richness with Practicality} \label{sec:practicality}

While multimodal sensing is widely advocated as one way to improve the accuracy and robustness of adaptive VR systems, the reviewed literature reveals a trade-off between sensing richness and practical deployability. First, in terms of data acquisition methods, simpler sensing configurations (e.g., wrist-worn EDA \cite{hadadi2024machine}, headbands \cite{liang2023eeg}, or skin-attached sensors \cite{i2018toward,bian2019design}) are more frequently adopted in interactive applications, whereas more complex setups (e.g., EEG combined with VR HMDs \cite{dey2019exploration}) are typically confined to controlled experimental environments. Therefore, we can see that modality selection could be influenced by practical constraints related to setup time, intrusiveness, and user burden. For example, wearable sensors offer relatively low setup cost \cite{fominykh2018conceptual}, while EEG-based devices often require additional preparation steps (e.g., electrode placement with conductive gel) \cite{feick2024predicting}, which can negatively impact user comfort and may introduce noise due to discomfort. For example, Islam et al. \cite{islam2021cybersense} chose not to use EEG due to its complexity and higher preparation time, despite the obvious correlation between EEG and cybersickness. This highlights a key gap between sensing capability and practical usability.

Moreover, another challenge in multimodal human-centered systems lies in data fusion and alignment across heterogeneous modalities, particularly when modalities differ in sensing mechanism or role. While existing multimodal fusion techniques are generally effective for integrating signals with similar temporal and semantic properties (e.g., biosignals, images, and audio \cite{islam2021cybersense,li2019gesture,han2021improving,zheng2022multimodal}), their applicability becomes less clear when modalities differ in structure and meaning. In particular, biosignals represent continuous physiological processes, whereas performance metrics (e.g., task completion time and accuracy) capture discrete outcomes. However, common standard data fusion approaches often assume commensurate representations in terms of sampling frequency, noise, and origin \cite{lahat2015multimodal}. Thus, this distinction places performance metrics at a higher level of abstraction, making direct integration with biosignals through standard data fusion techniques conceptually problematic. Notably, although performance measurements are common in adaptive human-centered systems, they are almost exclusively used for post hoc evaluation rather than as inputs to the adaptive loop.

As a result, current systems often isolate multimodal inputs architecturally rather than learning joint representations. For example, physiological data might estimate real-time states \cite{marinho2024eyes,feick2023investigating}, while performance metrics inform policy updates \cite{dey2019exploration, recenti2021toward,grootjen2024your} or offline evaluations \cite{delahaye2023avatar,dubovi2022cognitive,calvert2020impact,jin2023development}. While simplifying implementation, this design pattern relies on pre-defined user studies and post hoc performance evaluations \cite{kundu2023litevr}. Because these systems only learn from limited experimental sessions, they fail to capture long-term user patterns, thereby hindering the development of self-improving VR frameworks \cite{yu2025effects,li2024exploring}.

Another solution to balance information richness and deployment practicality is cross-modal learning \cite{kim2019deep,li2023deep}. Prediction models rely on content-aware inputs to ensure accuracy and generalizability. Although physiological data reflect individual differences in feedback, the high cost and complex device setup make it difficult to move from laboratory settings to consumer-grade devices \cite{liu2023emotionkd}. In contrast, lightweight models based on video often exhibit limited generalization performance due to a lack of awareness of users' physiological characteristics \cite{zhu2025towards}. To overcome these issues, researchers adopt cross-modal learning to extract personalized representations from physiological signals during the learning stage, which are then used to supervise and align the video encoder \cite{wang2025cross}. During the prediction stage, the model can perform subject-aware prediction using only visual stimuli. In this way, cross-modal learning reduces the reliance of prediction models on real-time physiological data and enhances the robustness of video-based prediction models. For instance, Zhu et al. \cite{zhu2025towards} proposed a cross-modal alignment framework in which physiological data features are learned during training. Then, during deployment, the system can run on consumer-grade hardware using only VR video input.

In summary, relying on these signals is rational given current technical constraints. However, this approach prioritizes robust measurement over how simple the system is to deploy or how smoothly users can experience it. In addition, while data fusion techniques have made substantial progress in integrating heterogeneous data streams, not all modalities integrate effectively. We identify three critical challenges: \textbf{CH1}: balancing data acquisition with UX, \textbf{CH2}: aligning and fusing diverse data streams, and \textbf{CH3}: ensuring data security.

\subsection{Integrating Reactive and Proactive Adaptation in VR} \label{sec:dis2}

One key trend emerging from recent studies is the increasing reliance on AI techniques to drive personalization in VR. Although AI techniques provide an innovative method to personalize content, reactive mechanisms have not yet been fully replaced by proactive ones because fully autonomous systems require immense computing power and can be unpredictable \cite{bellucci2025immersive}. Instead, the field is adopting a hybrid architecture. The following subsections explain this adoption in detail.

\subsubsection{From Reactive to Proactive} \label{sec:proactive}

Over the past decade, VR research has shifted away from passive, non-adaptive experiences to closed-loop systems that respond to real-time user data and environmental changes. According to the timing of adjustment, we could broadly categorize systems into two parts: reactive adaptation and proactive adaptation. Reactive adaptation follows a response-based mechanism. The system monitors user input, system state, or environmental changes, then takes corresponding action after a deviation is detected \cite{golpayegani2024adaptation}. In contrast, proactive adaptation follows a different logic: it predicts future conditions, user needs, or possible issues and makes adjustments before problems arise \cite{golpayegani2024adaptation}. The distinction is that reactive systems act after a change, while proactive systems act before one is needed.

\begin{figure}
    \centering
    \includegraphics[width=\linewidth]{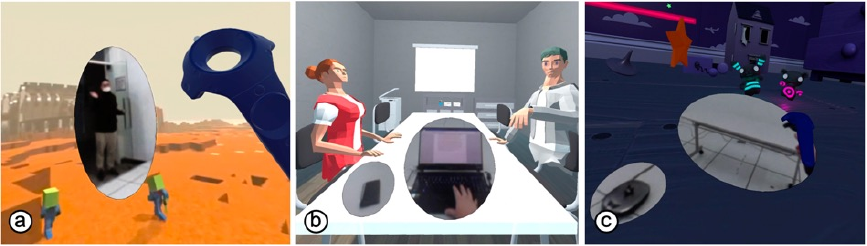}
    \caption{A user interface, RealityLens, that allows users to (a) communicate, (b) interact, and (c) avoid obstacles in the physical world in VR. This interface is customizable in terms of size and placement. From \cite{wang2022realitylens}.}
    \label{fig:realitylens}
\end{figure}

Across the reviewed literature, reactive adaptation remains the dominant paradigm. This is particularly common in applications such as gaming and immersive interaction, where system adjustments (e.g., difficulty adjustment or visual augmentation) are triggered by measurable signals (e.g., performance or behavioral input) \cite{marinho2024eyes,wang2022realitylens}, see Figures \ref{fig:realitylens}(a,b,c) as an example. This suggests that reactive approaches are preferred because they are conceptually straightforward and rely on observable indicators, making them easier to implement and assess \cite{marinho2024eyes}. However, this reactive nature means adaptations occur only after an event has emerged, which may disrupt UX in highly immersive or time-sensitive scenarios.

By contrast, proactive adaptation remains comparatively underexplored, despite its theoretical advantages in maintaining an immersive experience \cite{kraus2020effects}. This gap exists because systems struggle to predict user states, intentions, and environmental changes in real-time \cite{lopes2025generative}. As a result, this restricts the system's effectiveness in complex or open VR environments.

Recent advances in generative AI and LLMs enable the rapid creation of novel content, prompting researchers to design more flexible reactive and proactive systems. For example, AI-driven agents can dynamically adjust instructional content or interaction strategies based on user behavior \cite{gao2025pervrml}. For example, stimulating active students' behavior continuously to foster a positive classroom atmosphere and correct inappropriate student behavior \cite{liu2024classmeta}. However, a closer examination of these systems reveals that they do not operate as purely proactive systems. Instead, they usually adopt a hybrid strategy, where proactive mechanisms (e.g., content generation \cite{min2024public}) are combined with reactive components (e.g., affect classification \cite{bekele2016multimodal,fominykh2018conceptual}, behavioral correction \cite{liu2024classmeta}). This suggests that reactive and proactive adaptation are not mutually exclusive, but rather complementary strategies operating at different functional levels.

Some studies report that proactive adaptation offers clear advantages because it can better maintain system quality and avoid the reliability issues inherent in post hoc corrections \cite{kraus2020effects,xiao2003quiet}. However, depending on the application scenario, the usefulness and suitability of reactive and proactive systems vary. Reactive approaches are more robust in scenarios where system states can be reliably measured and validated \cite{wang2021enhanced}, whereas proactive approaches are more valuable in maintaining continuity in immersive or anticipatory interactions \cite{liu2024classmeta}. However, the current literature provides relatively few systematic comparisons of these strategies across application domains, particularly in terms of UX, system reliability, and computational cost across domains. With the rise of generative AI, additional design tensions have emerged, further complicating this situation. Sections \ref{sec:ai} and \ref{sec:domain} discuss how generative AI influences adaptive systems and explore the best approaches for incorporating generative content across application scenarios.

\subsubsection{The Rise of Generative Synthesis in VR} \label{sec:ai}

One key trend emerging from recent studies is the increasing reliance on  ML, affective computing, and more recently, generative AI to drive personalization in VR. The role of these techniques differs substantially between reactive and proactive approaches. Across studies, ML in reactive systems is mainly used for real-time state classification. These systems map sensor inputs (e.g., EDA, EEG, gaze) to discrete user states such as cognitive load \cite{dubovi2022cognitive,chiossi2023adapting,chiossi2024optimizing}, engagement \cite{lahiri2012design,porssut2021adapting,besga2025task}, or cybersickness \cite{uyan2024cdms,sameri2024physiology}. The resulting predictions are then passed to rule-based systems that trigger predefined adaptations, such as adjusting UI settings \cite{wu2024fanpad,wu2024gazefed} or altering the complexity of the virtual environment \cite{chiossi2023adapting}. Although this design pattern is effective for responsiveness and interpretability, it limits adaptability due to predefined rules and lacks the ability to generalize beyond preset scenarios.

On the other hand, proactive systems rely on predictive modeling algorithms capable of forecasting future states. Temporal models such as recurrent neural networks and long short-term memory networks are frequently employed to discover sequential dependencies in biosignals \cite{kundu2023litevr,tasnim2024investigating} and behavioral data \cite{azim2025your}. These approaches enable predicting states such as cybersickness \cite{kundu2023litevr,tasnim2024investigating}, navigation intention \cite{alghofaili2019lost}, or user actions \cite{yu2025effects}. However, these predictive capabilities are limited to short-term horizons and narrowly defined tasks, indicating that proactive adaptation remains constrained by model performance.

Recently, emerging technologies like generative AI and LLMs are beginning to enable more natural and context-aware interactions by powering adaptive virtual agents that can generate realistic dialogue and suggestions in real-time \cite{liu2024classmeta,li2024exploring,min2024public,wan2024building,ashby2023personalized,gao2025pervrml}. This enables more flexible and context-aware adaptation, particularly in applications such as education and social VR. For instance, LLM-powered agents can personalize explanations \cite{gao2025pervrml}, guide user interaction \cite{liu2024classmeta}, or mediate social dynamics through natural language and multimodal input, such as speech \cite{bellucci2025immersive} and gestures \cite{hu2025gesprompt}. Therefore, more studies employ a hybrid approach where generative AI handles high-level interaction and content creation, while lower-level state estimation remains dependent on ML models \cite{ojha2024dynamic}. However, the system must accurately interpret the user's internal state for effective adaptation. Such underlying state estimation relies on ML models that possess a ``black-box'' nature \cite{kundu2023litevr}. While the relationship between state estimators and subjective experience remains unclear, adding generative components further complicates the interpretation issues. Although explainable AI techniques (e.g., Shapley additive explanations) are increasingly adopted to improve transparency and enhance understanding, their use is often limited to post hoc analysis rather than being integrated into real-time decision-making \cite{raza2024optimized,kundu2023litevr,sameri2024physiology}. Nonetheless, such situations indicate the need for unified design frameworks to combine generative and predictive components within adaptive systems. As such, fully autonomous VR systems are less common.

To sum up, these findings suggest that while generative AI enhances VR adaptation flexibility, the lack of transparency in state estimation and content generation, and the heightened computational demand remain bottlenecks \cite{ribeiro2023virtual}. These problems make it difficult to determine how to combine different forms of intelligence for effective VR system adaptation.

\subsubsection{Domain-Specific Adaptation} \label{sec:domain}

\begin{figure}
    \centering
    \includegraphics[width=0.6\linewidth]{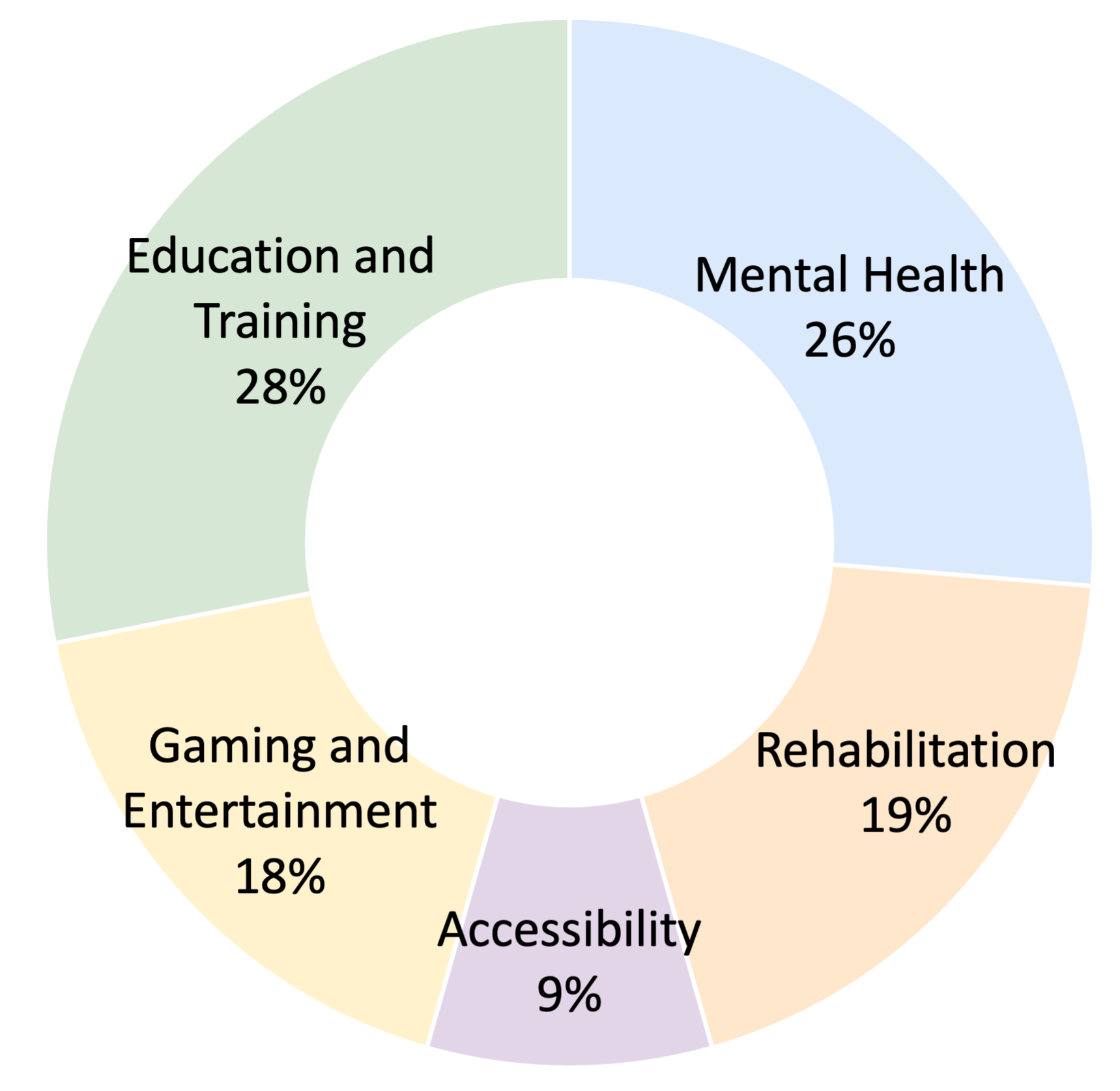}
    \caption{Percentage distribution of personalized VR systems across various application domains in the reviewed literature.}
    \label{fig:result}
\end{figure}

To better understand how different adaptive strategies are applied in practice, we demonstrate examples in three areas: gaming, education, and healthcare, to reveal how they leverage different strategies to achieve expected functions. Figure \ref{fig:result} illustrates the distribution of personalized VR systems across the reviewed studies' application domains.

The VR gaming industry has widely adopted personalization. Traditional gaming interaction is restricted to discrete inputs via peripheral devices (e.g., keyboard, mouse). VR enables embodied interaction by directly mapping the user's physical movements into the virtual space, providing richer entertainment and a more profound sense of presence \cite{tian2022review}. Early work primarily focused on reactive adaptation, such as dynamic difficulty adjustment based on player performance \cite{de2022adaptive,johnson2014knn,li2023design}, aiming to maintain engagement and prevent frustration \cite{brambilla2023tuning}. Recent work incorporates predictive and generative techniques to enable proactive personalization. For example, Ojha et al. \cite{ojha2024dynamic} leveraged VR's immersive affordances and AI-driven player modeling to develop an adaptive horror game system, which dynamically adjusts terrifying stimuli to individual players' specific fears and states. This reflects a shift from simple rule-based reactive adjustment toward combining reactive control with proactive content creation to deliver a more individualized experience. However, the immersive nature of VR inevitably introduces cybersickness \cite{heo2020eeg}. Also, the increased complexity of 3D interactions and virtual environments leads to more sophisticated systems and high-dimensional data streams.

Educational learning and professional training also benefit from adaptive VR systems. The virtual environment provides high fidelity and allows an embodied learning experience that traditional screen-based platforms cannot replicate \cite{dubovi2022cognitive}. It also provides risk-free simulations, avoiding hazardous situations that arise in scenarios and environments, such as laboratories and factories, in traditional in-person instruction \cite{chan2023study,zawadzki2020employee}. These systems predominantly utilize reactive adaptation to tailor the learning experience in real-time. As such, these platforms adjust learning content and task difficulty by monitoring task performance, engagement levels, and cognitive load \cite{obourdin2024unlocking,chan2023study}. While these adaptive systems offer great flexibility, teaching requires precision. Consequently, the integration of generative components is constrained to controlled applications, such as using an AI-powered chatbot to answer questions and LLMs to drive content generation \cite{song2024developing,bellucci2025immersive,gao2025pervrml}. Unlike VR gaming, which operates under more relaxed constraints and can include more proactive elements, educational systems must align strictly with established teaching frameworks to ensure accuracy \cite{pedram2023toward}. As a result, reactive strategies still dominate in VR training systems, as they ensure predictable performance and provide both instructors and students with greater transparency. This illustrates a trade-off between adaptability and reliability in such a domain. While proactive components offer open-ended exploration, structured reactive frameworks are necessary to ensure educational accuracy. Thus, proactive elements are typically involved in higher-level personalization rather than core instructional components in this domain.

Healthcare, including mental health, accessibility, and rehabilitation, represents one of the largest application areas, see Figure \ref{fig:result}. Traditional digital health platforms rely on passive screen-based interaction. In contrast, VR healthcare applications leverage embodied interaction and sensory immersion to facilitate treatments like exposure therapy \cite{mahmoudi2021automated} and motor rehabilitation \cite{bouatrous2023new} by mapping the user's physical movement into a 3D environment~\cite{wang2023movement}. While traditional in-person therapy remains the clinical standard, VR offers unique advantages by providing controllable and reproducible environments that are often impossible to achieve in physical settings \cite{makinen2022user}. For example, it could be difficult and risky when exposing a patient to specific triggers in in-person therapy, but VR provides a relatively safer approach since stimulus intensity can be modulated \cite{liu2022virtual,Wang2024modulation,zhang2024touchmark}. As we discussed earlier, gaming and learning (training) reveal the trend of incorporating proactive functions to some extent, but this trend is relatively rare in the healthcare sector since clinical safety and precision are prioritized. An emerging trend is the use of ML algorithms to analyze vast datasets from wearables and remote monitoring devices to predict potential health issues \cite{bian2019design,kritikos2021personalized,bouatrous2023new}. While predictive modeling and proactive interventions (e.g., early detection of health issues and therapy planning \cite{lee2022towards}) hold potential, their adoption is limited by the high stakes of clinical decision-making and the complexity of modeling long-term health trajectories. VR technology provides a technical foundation for emerging concepts such as digital twins, enabling more precise clinical treatment. However, such approaches remain largely exploratory due to technical and ethical challenges \cite{katsoulakis2024digital}. As a result, proactive techniques in healthcare often act in supportive roles rather than driving core clinical decisions \cite{li2024exploring}.

A cross-domain comparison reveals that the optimal balance between reactive and proactive adaptation depends on a holistic assessment of safety and the degree of interaction freedom. At one end, we see VR gaming systems that leverage VR's embodiment and interaction abilities to prioritize innovation \cite{tian2022review}. Conversely, healthcare VR remains more restricted. The risk of triggering adverse side effects causes systems to choose transparent and reactive frameworks to ensure medical safety \cite{liu2022virtual}. Educational systems fall between these extremes, showing a hybrid approach that utilizes LLMs for innovation and more personalized content while maintaining established educational standards \cite{pedram2023toward}. Despite the potential of these approaches, a critical gap remains in understanding how domain-specific constraints influence reactive control and proactive autonomy. We summarized these challenges as follows: \textbf{CH4}: coordinating practical deployment for real-world settings, \textbf{CH5}: optimizing computational efficiency for real-time adaptation, and \textbf{CH6}: enhancing transparency of adaptive decisions.

\subsection{Impact on User Experience} \label{sec:dis3}

While many studies demonstrate promising outcomes in controlled environments, transferring these results into daily applications remains limited \cite{xie2021review}. In the following, we analyze these challenges from multiple aspects to bridge the gap between experimental innovation and practical application.

\subsubsection{Limitations of Performance Metrics}

Current studies often rely on performance metrics, self-report questionnaires, and physiological signals to assess the effectiveness of adaptive models and mechanisms, as mentioned in Section \ref{sec:rq3}. While such measures can indicate functional efficiency, such as higher task scores, fewer errors, or stable cognitive load, they may provide incomplete information about UX during virtual immersion \cite{sachete2025beyond}. As pointed out in Kosch et al. \cite{kosch2023survey}'s work, frequently used HCI questionnaires (e.g., NASA-TLX for cognitive load measurement) are adopted from other fields without validation. Similarly, from Table \ref{tab:questionnaire}, we see that similar questionnaires are also widely used in VR studies to assess pre- and post-immersion performance. These questionnaires are not specifically designed to account for the high-dimensional, embodied interactions inherent in VR. Although there are VR-specific questionnaires (e.g., Virtual Reality Neuroscience Questionnaire and Virtual Experience Questionnaire for measuring VR experience), others fail to capture factors arising from VR immersion.

In particular, performance metrics are frequently used as primary indicators for task-driven applications such as VR training \cite{lin2022study}. Improvements in accuracy, efficiency, or error reduction are often interpreted as signs of successful improvement \cite{dey2019exploration}. However, this conclusion overlooks the fact that performance gains do not necessarily reflect positive UX outcomes. For example, adaptive systems may reduce task difficulty or constrain user behavior to optimize outcomes, but hide underlying negative experiences such as frustration and anxiety. This suggests that improvement may come at the expense of user engagement. A similar issue arises with subjective and affective measurement. Although questionnaires such as System Usability Scale, NASA-TLX, and Game Experience Questionnaire are commonly used to capture user perceptions, they are often treated as secondary metrics and reported descriptively rather than integrated into the evaluation of adaptive mechanisms \cite{zhao2023adaptive,li2024exploring}. As a result, there is limited systematic analysis of how adaptive strategies impact subjective experience, particularly in relation to negative affect or long-term user satisfaction.

Therefore, these findings highlight misalignment between measurable efficiency and actual UX. Relying on an incomplete evaluation framework may create an optimization paradox in which a system helps users achieve good task performance at the expense of user motivation or triggers frustration. The current focus on functional gains neglects the user’s psychological cost. To address the gaps mentioned in Section \ref{sec:dis3}, we identify \textbf{CH7}: transitioning from static post hoc evaluation to continuous, multidimensional assessment.

\subsubsection{Well-Being and Long-term Engagement}

Long-term efficacy is critical to consider since it influences performance retention and user re-engagement. Especially in training and rehabilitation scenarios, sustained improvement that occurs through recurring use serves as a vital indicator for performance. While short-term performance gains are commonly reported (e.g., increased typing efficiency \cite{wu2024fanpad} and enhanced cognitive function through repeated VR therapies \cite{zhai2021virtual}), the reviewed literature reveals a clear lack of longitudinal evaluation, with only a limited number of studies assessing system effectiveness over extended periods \cite{ip2018enhance,du2024lightsword,yu2025effects,bellucci2025immersive}. The lack of longitudinal designs makes it difficult to determine whether these improvements reflect genuine skill acquisition, sustained behavioral change, or temporary motivational boosts, weakening the robustness of the adaptive system. This is caused by challenges in deploying VR systems under controlled research settings \cite{azim2025your}. In non-VR studies, participants can wear portable sensors to acquire data and monitor performance for weeks without disrupting their daily routines. In contrast, VR hardware creates a unique barrier for longitudinal data collection. Because VR HMDs are bulky and restrictive, VR experiments need to be conducted under controlled laboratory settings to ensure user safety. This makes it more difficult to observe how adaptive systems perform in the wild over extended periods. As a result, current studies rely more on analyzing short-term effectiveness and leave long-term impact underexplored.

While we may assume that greater immersion and personalization inherently lead to better learning outcomes, existing evidence also suggests that short-term gains may not translate into long-term improvement \cite{xie2021review}. For example, Juliano et al. \cite{juliano2022increased} reported that training on motor skills in an immersive VR environment resulted in lower long-term retention and context transfer than training on a computer. Therefore, it raises critical questions about how adaptive mechanisms in VR may inadvertently hinder skill formation. As a result, the effectiveness of VR training and relevant personalized adaptation remains open to further validation in future research. It also raises a broader concern about user well-being in the long term. Prolonged interaction with adaptive systems may introduce unexpected effects such as over-reliance on system adjustment, which is rarely evaluated in current studies \cite{hoover2021designing}.

Taken together, these observations highlight the lack of temporal dynamics in VR adaptation. Future research should move beyond short-term evaluation and incorporate longitudinal study designs that capture retention, skill transfer, and sustained engagement \cite{zhai2021virtual,du2024lightsword}. Establishing an evaluation framework that accounts for possible negative long-term effects is essential to validating the feasibility of adaptive VR systems.

\section{Challenges and Future Research Directions} \label{sec:future}

\begin{figure*}[htbp]
    \centering
    \includegraphics[width=0.9\linewidth]{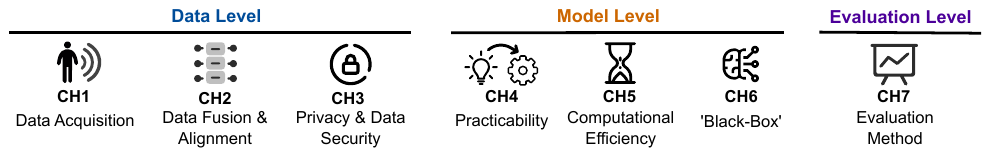}
    \caption{An overview of current challenges categorized into data, model, and evaluation levels.}
    \label{fig:challenges}
\end{figure*}

During the review process, we identified some areas that need attention in future research. Based on the discussion in Section \ref{sec:discussion}, we outline the challenges \textbf{CH1-CH7} the field currently faces in Figure \ref{fig:challenges}. We highlight the following research directions that could address or ease the aforementioned challenges.

Existing systems show a clear trend toward incorporating multimodal data to improve the accuracy and robustness of user state estimation. However, as discussed in Section \ref{sec:practicality}, this trend introduces a fundamental trade-off between sensing richness and practical usability.

\textbf{CH1.} Current VR adaptive systems face sensor overload, requiring users to wear an HMD with multiple sensors to capture physiological signals. This setup significantly increases setup complexity and reduces users' wearing comfort. Also, it causes fatigue and reduces the user's sense of presence due to equipment weight and tactile disturbance. Although high-quality data are collected, UX is diminished, which could affect participant performance during studies. Therefore, a key research direction is understanding which modalities are necessary and sufficient for specific tasks. This requires a more precise mapping between sensing modalities, application domains, and user states, enabling minimal and effective sensing structures.

\textbf{CH2.} In addition to data acquisition challenges, the integration of heterogeneous data remains another issue. As highlighted in Section \ref{sec:practicality}, existing fusion techniques are often applied without fully accounting for differences in temporal resolution, semantic meaning, and abstraction level (e.g., physiological signals vs. performance metrics). In addition, physiological and behavioral signals are highly context-dependent in VR. The same signal patterns can mean different things depending on factors such as the virtual environment, interaction technique, and immersion level. This results in models that are difficult to interpret and generalize. Future research should move toward more principled fusion frameworks, such as attention-based architectures and temporally adaptive models, that account for modality heterogeneity \cite{liang2021attention,wu2017event,sheng2015adaptive}. VR systems may benefit from event-based synchronization based on interaction events (e.g., object manipulation and locomotion transitions) instead of relying solely on global timestamps.

\textbf{CH3.} Privacy is another issue that cannot be ignored, because VR adaptive systems increasingly rely on biosignal data, and these data carry personal information about subjects. From a technical perspective, federated learning could be applied \cite{wu2023privacy}. Through training on private data at the local server and only sending model updates to the global model, it reduces the risk of leaking sensitive information from raw data. Also, cross-modal learning mentioned in Section \ref{sec:practicality} is another possible solution to ease privacy issues. In general, privacy considerations must be built into a system's architecture at the start instead of being added as an add-on. This approach is supported by emerging industry standards and frameworks, such as ISO/IEC 27701 for managing personal data \cite{folorunso2024impact}.

Beyond challenges in data level, our review identifies challenges related to the deployment of adaptive models in real-world settings.

\textbf{CH4. }In terms of challenges at the model level, most current systems are used within controlled laboratory environments. As the user community expands, systems must scale to support real-world deployment. This reflects a gap between experimental validation and practical deployment. There is a fundamental difference regarding where and how adaptive VR is used. VR gaming mirrors non-VR gaming, where users purchase and play in home environments. In contrast, healthcare VR remains limited to professional clinical settings. Much like traditional therapy, VR therapy settings could ensure sufficient hardware requirements. Thus, deployment issues may be more concerning in sectors like gaming and education. As we mentioned earlier, adaptive systems rely on external biosensors to detect the user's current state in order to implement personalized adjustments \cite{hadadi2022prediction}. In labs, researchers can equip users with various devices, but home users typically do not have them. To address this, future development should focus on integrating sensing into VR devices. In addition, adaptive models must be optimized to operate with a minimal yet effective sensing structure rather than adding multiple signals \cite{pei2024eeg}. Addressing this requires both technical solutions and system-level considerations.

\textbf{CH5.} Moreover, system responsiveness remains a problem during practical deployment. Nowadays, systems use high-frequency biometric or behavioral data as input for sophisticated models, introducing significant computational demands that can limit real-time interaction \cite{pei2024eeg}. Computational efficiency requirements are stricter in VR than in conventional HCI systems. VR is highly sensitive to temporal discontinuities, so any interruption can break immersion and affect the overall UX \cite{tao2022integrating}. While techniques such as model compression and knowledge distillation offer potential solutions \cite{kuzmin2023pruning,he2022knowledge}, the underlying challenge is the lack of cooperative design among sensing, modeling, and systems, which is vital for efficient adaptive VR systems. Additionally, using essential modalities, as mentioned earlier when tackling \textbf{CH4}, could also help reduce computational pressure.

\textbf{CH6. }Furthermore, the ``black-box'' problem of AI models, without revealing the internal logic or reasoning, the lack of transparency raises concerns about bias, accountability, and controllability. Although explainable AI techniques (e.g., SHAP \cite{lundberg2017unified}) are increasingly adopted, they are often applied post hoc, and this method alone cannot fully resolve the challenge or fully address the need for real-time transparency. \cite{pavel2025patchfusionvr}. Such non-transparent decisions also affect user trust. Therefore, giving users some degree of control over the adaptive strength and acknowledgment of model decisions may improve transparency. Thus, research needs to determine how to present explanations in accessible formats for all users.

Finally, the review highlights limitations in how adaptive VR systems are evaluated.

\textbf{CH7. }As discussed in Section \ref{sec:dis3}, diverse assessment methods are used in current research. Although multiple factors shape UX during immersion, their individual contributions to UX remain poorly understood \cite{hameed2024authenticity}. Thus, a method that could help researchers determine how each factor contributes to overall UX during VR immersion would be effective for future system design. We advocate for the development and adoption of standardized protocols and validated instruments to assess the effectiveness of personalization strategies. In addition, current evaluation practices focus on short-term performance metrics, with limited consideration of long-term engagement, learning retention, and user well-being due to VR setup challenges. This is especially problematic in domains such as rehabilitation, education, and mental health, where monitoring long-term engagement and improvement is crucial \cite{juliano2022increased}. Future studies should focus more on examining how personalized systems influence user adaptation, learning retention and transfer, and sustained engagement over time \cite{makarova2023virtual}. One promising direction is the adoption of multi-phase study designs, in which adaptive VR systems are evaluated across different phases, enabling researchers to distinguish novelty effects from sustained learning and to observe how adaptation strategies influence users over time. Beyond subjective metrics, there is an absence of evaluation datasets to compare systems objectively. Considering the diversity of applications, domain-specific evaluation datasets are crucial for consistent evaluation within comparable contexts. However, VR datasets are often difficult to reuse or share due to challenges such as varied experimental setups and privacy concerns. In contrast to AI research, benchmark datasets such as ImageNet \cite{deng2009imagenet} and COCO \cite{lin2014microsoft} have enabled rapid progress. While some efforts exist in VR, such as VREED \cite{tabbaa2021vreed}, they lack the scale and generalizability of AI benchmarks. Therefore, promoting the creation and sharing of standardized datasets is essential to advancing the training and evaluation of personalized VR systems \cite{wang2023ieee,otto2019virtual}.

\section{Conclusion} \label{sec:conclusion}

In this paper, we presented a comprehensive literature review of adaptive VR systems over the last decade. Three research questions are provided to guide the review process. We categorized the workflow of adaptive VR systems into five stages, analyzing the current focus and summarizing emerging trends in each stage. The aim of this survey is to understand how adaptive VR systems progress, especially with the maturation of multimodal sensing and the advancement of AI models. Subsequently, we discussed the underlying causes of certain trends and explored future development trajectories. Our review highlighted a growing body of work that uses multimodal data as input to refine system performance, employs ML models in model-driven inference systems, and utilized LLMs for generative adaptation. We also identified a shift from simple reactive to hybrid adaptation, which combines the strengths of reactive and proactive strategies, yielding more intelligent adaptation for education and entertainment. Whereas the healthcare domain relies more on reactive systems due to requirements for safety and controllability, it leverages AI algorithms to improve accuracy in user state identification. Finally, key challenges include the trade-off between data richness and practicality, transparency in AI-driven adaptation, computational demand, fragmented evaluation practices, and insufficient longitudinal validation. By identifying current trends, limitations, and emerging opportunities across data, strategies, and impact, this work aims to guide future research toward more effective adaptive VR systems.

\section*{Author Contributions}

Tangyao Li: Conceptualization, Methodology, Investigation, Writing - original draft, Writing - review \& editing, Validation, Visualization. Yitong Zhu: Investigation, Writing - review \& editing. Hai-Ning Liang: Investigation, Writing - review \& editing. Yuyang Wang: Conceptualization, Investigation, Methodology, Supervision, Writing - review \& editing, Project administration, Funding acquisition.

\section*{Disclosure Statement}
The authors declare no conflict of interest.

\section*{Funding}
This work was supported by Guangdong Basic and Applied Basic Research Foundation under Grant No. 2025A1515110098.

\section*{Data Availability Statement}
No new data were collected or generated for this study.


\bibliographystyle{ACM-Reference-Format}
\bibliography{reference}










\end{document}